\documentclass[showpacs,preprintnumbers,showkeys,twocolumn,prl,floatfix]{revtex4}
\usepackage{epsfig}

\begin{document}

\count60 = \time

\count62 = \count60

\divide\count60 by 60	

\count61 = \count60

\multiply \count61 by -60
\advance \count61 by \count62

\ifnum \count60>12 \global\advance \count60 by -12\fi

\divide\count62 by 60	

\def\timetoday{\ifcase\month\or
  January\or February\or March\or April\or May\or June\or
  July\or August\or September\or October\or November\or December\fi
  \space\number\day, \number\year, 
  \number\count60:\ifnum \count61<10 0\fi\number\count61
  ~\ifnum \count62>12 pm\else am\fi}

\title{Meissner Effects and Constraints}
\date{\timetoday}

\author{James P. Sethna}
\affiliation{
Laboratory of Applied Physics, Technical University of Denmark, %
DK-2800 Lyngby, DENMARK, and NORDITA, DK-2100 Copenhagen \O, DENMARK
and
Laboratory of Atomic and Solid State Physics (LASSP), Clark Hall,
Cornell University, Ithaca, NY 14853-2501, USA}
\author{Ming Huang}
\affiliation{Graduate School of Business, Stanford University,
Stanford, CA 94305-5015}

\begin{abstract}
We notice some beautiful geometrical defects found in liquid crystals,
and explain them by imposing a constraint.  We study the way constraints
can occur, and introduce the concept of massive fields.  We develop the
theory of magnetic field expulsion in superconductors as an example.
We notice strong analogies with the formation of grain boundaries in
crystals, and realize that we do not understand crystals very deeply.

\noindent
{\bf 1991 Lectures in Complex Systems,} Eds.\ L.~Nagel and D.~Stein,
Santa Fe Institute Studies in the Sciences of Complexity, Proc.\ Vol.\ XV, 
Addison-Wesley, 1992.
\end{abstract}

\maketitle

In the last lecture, I explained how condensed--matter and high--energy
physicists used topological theories to describe defects excitations in
solids.  In this lecture, I'm going to make fun of 
topology.\footnote{Everything I know about focal conics and
smectic liquid crystals\cite{Kleman} was explained to me by Maurice 
Kl\'eman, who also was one of the originators of the topological theory 
of defects.  No disrespect is intended.}  Actually, I'm going to
start by talking about constraints, then ``massive'' fields and
how they produce constraints.  I'll then turn to the Meissner--Higgs
effect in superconductors, and finally explain why I don't understand
crystals.

\section{Constraints}

\begin{figure}[thb]
\epsfxsize=2.5truein
\epsffile{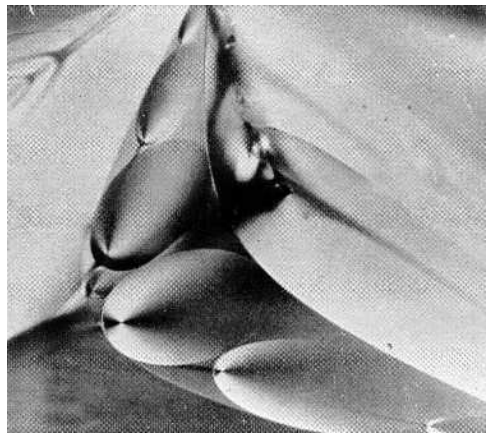}
\caption{Ellipses: Defects in a Liquid Crystal.
This is a drop of smectic A liquid crystal, squeezed between two
microscope slides.  The microscope is focused on the surface
of the drop, where it contacts the glass.  Notice the beautiful,
geometrical ellipses.  Notice that a line seems to exit from the focus
of each ellipse.  This line turns out to be a hyperbola (figure~4).
The visible ellipses and the hyperbolas are where the smectic layers
pinch off to form cusps.  These defects are {\it not} topological:
they are geometrical consequences of the constraint of equal layer
spacing.  From \cite{deGennes}, figure 7.2, photo by C.~Williams.
}
\label{fig:FocalConics}
\end{figure}

Consider figure~1.  See the beautiful ellipses and hyperbolas?  Remember 
that topology treats ellipses as rubber bands.  Any topological theory has 
got to miss the key feature of the  beautiful structures produced here: the
geometrically perfect ellipses with dark lines coming out of one focus.

Figure~1 is a photograph of a drop of fluid, squeezed between two microscope
slides.  The microscope is focused, let's say, on the surface between the
fluid and the bottom microscope slide: the ellipses are stuck onto the
glass.  The sizes of the ellipses are roughly given by the thickness
of the fluid layer.  The fluid is a smectic~A liquid crystal.  
deGennes\cite{deGennes} has a fine discussion and some nice pictures too.

In 1910, Friedel figured out why this liquid crystal forms these geometrical
structures.  He learned all he needed to know from his high--school
geometry class.  He actually worked backward, and used the ellipses
to deduce what kind of broken symmetry the liquid had.  Since none of you
were taught about the cyclides of Dupin in high school\footnote{Bertrand
Fourcade tells me that even the French stopped teaching them.},
I'd better start with the broken symmetry and work forward.

\begin{figure}[thb]
\epsfxsize=2.5truein
\epsffile{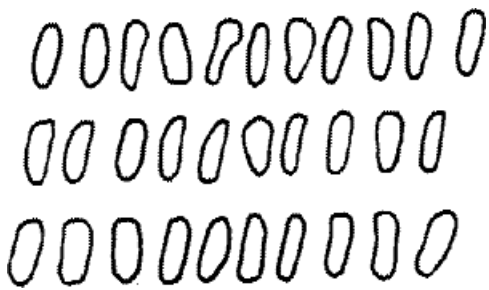}
\caption{
{\bf Order in Smectic Liquid Crystals.}
Smectic liquid crystals are formed of layers of molecules.  In each
layer, the molecules are in a random, liquid configuration.  Crystals
have broken translational symmetry along three independent axes:
smectic~A liquid crystals have broken translational symmetry in
only one direction (normal to the layers).
}
\label{fig:SmecticSchematic}
\end{figure}

Smectic liquids form equally spaced layers.  Some of them are compounds
that, like soap, naturally form membranes and films: I think smectic is the 
Greek word for soap.  Others are long thin molecules like nematics,
which for some reason not only line up, but segregate into planes
(figure~2).  The molecules have liquid--like order in the planes.  
Like crystals, they have a broken translational symmetry, but only 
in one of the three directions.

\begin{figure}[thb]
\epsfxsize=2.5truein
\epsffile{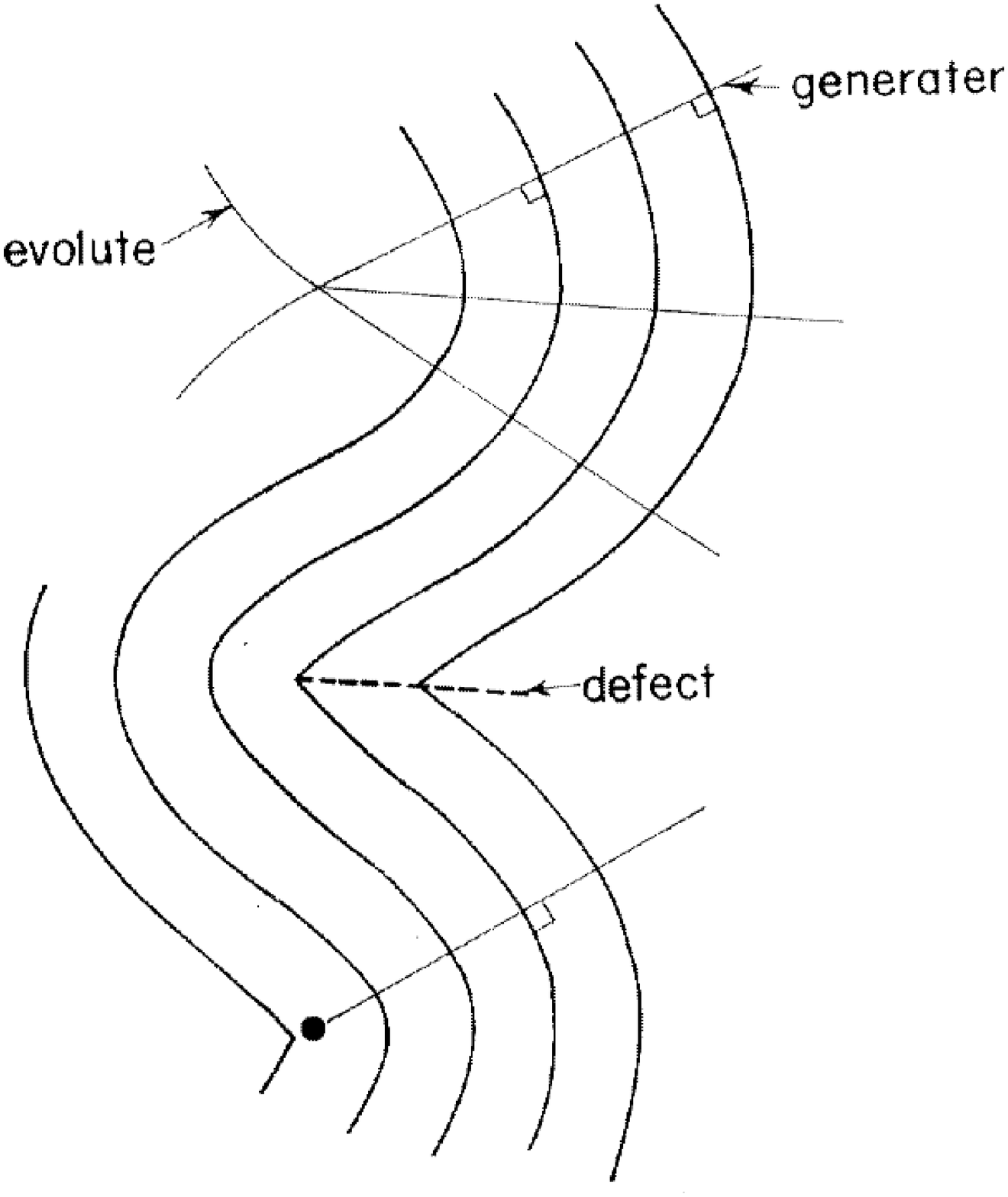}
\caption{
{\bf Equally Spaced Layers: Defect Formation.}
Here we consider a two-dimensional analogue of a smectic liquid crystal.
The smectic layers are represented by curves in the plane.  The lowest
energy state, of course, consists of parallel straight layers, but
the layers often settle into more complicated patterns, with
defects.  For reasons that we discuss in this lecture, and which
are not completely understood, smectic layers will deform by bending,
but will remain strictly equally spaced (except very near boundaries and
defects).\hfil\break
The constraint of equal layer spacing has weird nonlocal consequences.
First, one can see that as one moves outward the concave regions become
more pinched, and eventually form cusps.  Second, one can see that
a line perpendicular to one layer (a generator) will be perpendicular to
the next one too.  These generators intersect on a surface known as the
{\it evolute}, and it is when the layers hit the evolute that a defect
occurs.  As one sees here, the defect is a line of pinched surfaces:
in three dimensions it is typically a two-dimensional surface.  This
costs lots of energy.  The only way in two dimensions to have a point--like 
low--energy defect is to have concentric circles: only circles have
zero--dimensional evolutes.  The only way in three dimensions to have
one--dimensional evolutes\cite{Hilbert} is to have cyclides of Dupin: 
the defects are ellipses and hyperbolas passing through one another's foci 
(figures~1 and~4).
}
\label{fig:Smectic2D_3}
\end{figure}

Now, the important excitations for smectics are those that bend the
layers.  In figure~3, we see a two-dimensional analogue of the 
smectic liquid crystals: equally spaced curves in the plane.
Suppose we start with one curve and work outward.  As you can see
from the figure, the next curve is not precisely the same shape:
keeping the surfaces at an equal spacing makes concave regions
become sharper and convex regions become more rounded.  It is easy
to see that eventually the concave regions will become pinched:
these pinches are the defects.  They are not topological defects,
since rounding them a bit makes them go away: they are geometrical
defects produced by the constraint of equal layer spacing.

\begin{figure}[thb]
\epsfxsize=2.5truein
\epsffile{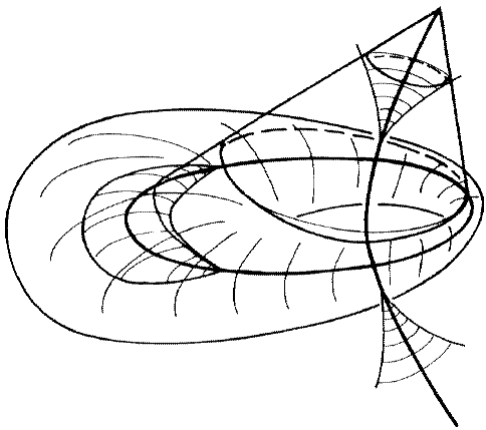}
\caption{{\bf Focal Conic Defect.}
Here we see the smectic surfaces which form the focal conic defects seen
in figure~1.  These are the cyclides of Dupin.  The surfaces go from
banana--shaped to squashed doughnuts to apple--shaped.  The points on
the bananas and the dimples at the stem and bottom of the apples are
defects, which scatter light and show up in figure~1.  (Only the dimples
of the apple are shown.)  The banana defects
lie on an ellipse, and the apple defects lie on a hyperbola which passes
through the focus of the ellipse.\hfill\break
Usually, the whole pattern isn't found in the experimental sample.  As 
you see in figure~1, the domains aggregate together in clumps.  Each
ellipse in figure~1 has a conical region for its smectic layers.
}
\label{fig:FocalConicSchematic}
\end{figure}

Most curves, like the one shown in figure~3, form one--dimensional
pinched regions: only concentric circles and structures made from
them can keep the pinched regions to points.  In three dimensions,
the only equally spaced surfaces with points as pinched regions are
concentric spheres.  Now, what Friedel knew and you don't know is that the
only 3-D surfaces with one--dimensional line-like defects are the 
cyclides of Dupin,\cite{Hilbert} {\it and the pinched regions form 
ellipses and hyperbolas}.\footnote{Actually, the canal surfaces also
have singularities confined to one--dimensional regions\cite{Kleman,Hilbert},
but let's not get bogged down.}  

\begin{figure}[thb]
\epsfxsize=2.5truein
\epsffile{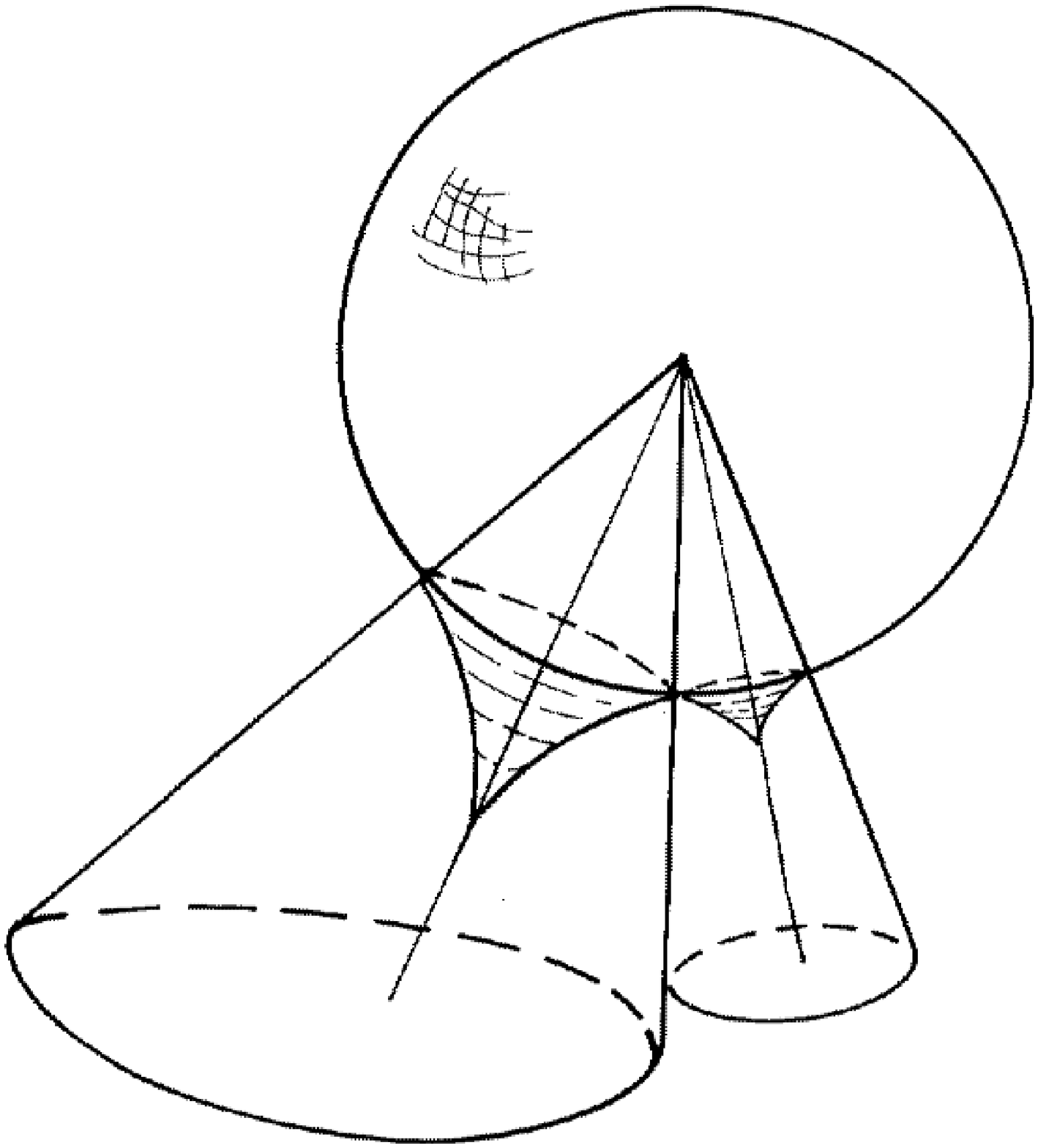}
\caption{{\bf Focal Conic Defect Meshing Onto Concentric Spheres.}
The conical regions in figure~4 combine into compound defects by meshing
onto the concentric sphere defect.  Concentric spheres are the only
surfaces with zero--dimensional defects.  The surfaces on the edges
of the cones mesh smoothly onto the concentric spheres.
}
\label{fig:ConcentricSpheres5}
\end{figure}

Figure~4 shows the cyclides of Dupin.  Notice that they pinch off on
two curves: an ellipse and a hyperbola.  The hyperbola is perpendicular
to the plane of the ellipse, and passes through its focus.  That's what
you see streaming out of the foci in the photo, and why you don't see
one for each focus.  My contribution to the field (with Maurice Kl\'eman)
was to realize that these cyclides of Dupin fit together nicely inside
concentric spheres, which explained neatly the ways the ellipses 
always seemed to fit together (figure~5).  Maybe the concentric spheres
form because the layers nucleate on a dust particle on one of the microscope
slides: when the spheres touch the other slide, the concentric spheres get
twisted (they like to sit perpendicular to the glass) and the ellipses
and hyperbolas form to relieve the strain.

Now, why do I show you this?  It isn't just to show that there is more
to the world than topology.  Mostly, it's to illustrate the two themes
of this lecture: constraints, and expulsion.

If we define an order parameter $\hat n$ for the smectic to be the
unit normal to the smectic layers ($\hat n^2 = 1$), then the constraint 
that the layers be equally spaced implies
\begin{equation}
\label{eq:curl}
{\rm curl}\, \hat n = \left(
	\matrix{\partial n_z/\partial y - \partial n_y/\partial z \cr
		\partial n_x/\partial z - \partial n_z/\partial x \cr
		\partial n_y/\partial x - \partial n_x/\partial y \cr}
		\right) = 0.
\end{equation}
(This is derived, for those who know a bit about vector calculus,
in the appendix.)  This is a remarkably powerful constraint.  For example,
knowing the position of one layer determines all the others!  We show
this mathematically in the appendix, but you saw it physically in
figure~3: given one layer, there is only one way to place the next one
preserving exactly equal spacing.

There is a pretty good analogy here to analytic continuation.  For
those of you who know about complex analysis, you know that an analytic
function obeys the Cauchy--Riemann equations.  If we let 
$n(x+i y) = n_x(x+iy) + i n_y(x+iy)$ be an analytic function, then 
\begin{equation}
\label{eq:analytic}
\left(\matrix{
		\partial n_x/\partial x - \partial n_y/\partial y \cr
		\partial n_x/\partial y + \partial n_y/\partial x \cr}
\right)=0.
\end{equation}
As you know, analytic functions have really bizarre properties.  If you
know an analytic function in a small region, you can figure it out
everywhere else, just like the order parameter in smectics.  The point 
singularities of
analytic functions have a rich and interesting classification (simple
poles, essential singularities, ...).  Both in analytic functions
and in our smectic problem, constraints on the derivatives of our
order parameters produced really bizarre, nonlocal, geometrical consequences.

\section{Massive Fields}

We've discovered that constraints can have beautiful, geometrical consequences.
How are the constraints enforced?  Clearly, it is possible to stretch the
smectic layers apart, or to compress them together: why doesn't this happen 
in practise, especially when the layers are being bent and twisted?
The curl of $\hat n$ is constrained to zero.
Why are magnetic fields pushed completely out of superconductors?
The magnetic field is constrained to zero.  Why isn't it possible to find 
an isolated quark in nature?  Quarks have non--zero ``color'', and 
the net color is constrained to zero.

These constraints come from minimizing the energy.  Saying that magnetic
fields can happen inside superconductors is just like saying that marbles
can sit on the side of a hill: it can happen, but not if the marbles
are allowed to roll to minimize their energy.  Under what conditions
does the energy enforce a constraint?  We say that it happens when the order
parameter field develops a {\it mass}.  We'll explain this term in a moment,
but let's first give a simple example.

Suppose we have a fluid in one dimension.  The density of a fluid is the 
important variable in describing its state.  Suppose the density of the fluid
is $\rho_0 + \rho(x)$, where $\rho_0$ is the ideal density (which the fluid
would have if left to itself) and the order parameter $\rho(x)$ describes
the deviation from the ideal density.  A sensible free energy might be
\begin{equation}
\label{eq:liquid}
{\cal E}_{fluid} 
	= \int dx\, (1/2) (d\rho/dx)^2 + (1/2) m \rho^2.
\end{equation}
The first term in the energy resists sudden changes in the density: having
a high density region right next to a low density region costs extra.
The second term in the energy says that deviations from the mean density
cost energy, with $m$ a coefficient which says how much deviations cost.
Unlike phonons, where the order parameter $u(x)$ could be uniformly
shifted without energy cost, here the lowest energy state happens when
the density is at its mean value $\rho(x)=0$.

What happens when we try to find the minimum energy state?  Clearly
the best we can get is the ideal state $\rho(x)\equiv 0$, which has
zero energy ${\cal E}_{fluid}$.  Perhaps, though, we're pulling on the density
at the two ends (figure~6).  If the liquid is in a trough of length $L$, we'll 
insist that $\rho(0) = \rho_i$ and $\rho(L) = \rho_f$.  What configuration
$\rho(x)$ minimizes the energy then?  Clearly, it should sag towards
$\rho_0$ inside, but how?

\begin{figure}[thb]
\epsfxsize=2.5truein
\epsffile{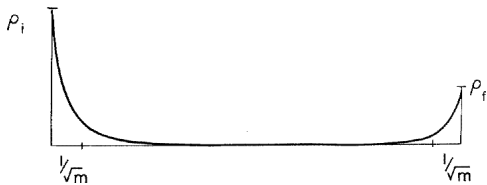}
\caption{{\bf Massive Fields Decay Exponentially.}
Minimizing the energy ${\cal E}_{fluid}$ in equation \ref{eq:liquid}, with
boundary conditions $\rho(0) = \rho_i$ and $\rho(L) = \rho_f$.  It is
easy to understand physically what is happening.  The system wants
to achieve $\rho = 0$, and it sags to that value as quickly as it
can, balancing the costs of $(d\rho/dx)^2$ energy against the gain.
The solution decays exponentially to zero with a decay constant
$\sqrt{m}$.
}
\label{fig:Massive}
\end{figure}

Here I'll show you a simple case of what's called the calculus of variations.
I apologize for the math, but it is really a useful method.
The trick is to realize that if $\rho(x)$ is the minimum energy configuration,
then $\rho(x) + \delta(x)$ must have a higher energy, whatever 
$\delta(x)$ we might choose.  
\begin{eqnarray}
\label{eq:vary}
{\cal E}(\rho + \delta) &-& {\cal E}(\rho) \\
	&=&\int ( d\rho/dx d\delta/dx + m \rho(x) \delta(x) \nonumber \\
	      &+& (1/2) (d\delta/dx)^2 + (1/2) m \delta^2)\,dx \nonumber \\
	&\ge& 0. \nonumber
\end{eqnarray}
Now, if we confine our attention to small $\delta(x)$, we can ignore
the last two terms (because they are quadratic, rather than linear, in
$\delta$).  The first term we integrate by parts, so 
\begin{equation}
\label{eq:parts}
\int_0^L dx\,  d\delta/dx\, d\rho/dx = (\delta\, d\rho/dx) \Bigl|_0^L\Bigr.
 - \int_0^L dx\, \delta\, d^2\rho/dx^2.
\end{equation}
Now, $\delta$ mustn't change the values at the endpoints, so 
$\delta(0) = \delta(L) = 0$ and the boundary terms in \ref{eq:parts} drop out.
We're left, then, with the equation
\begin{equation}
\label{eq:dE}
{\cal E}(\rho + \delta) - {\cal E}(\rho) \approx
	\int dx\, (-d^2\rho/dx^2 + m \rho(x)) \delta(x) \ge 0.
\end{equation}
Now, this must be true for any $\delta(x)$ we choose.  This can only
happen if $-d^2\rho/dx^2 + m \rho(x) = 0$, so $\rho''=m\rho$.  

The solutions to this equation are, of course, 
$\rho = A e^{-\sqrt{m} x} + B e^{\sqrt{m} x}$.  We can vary the arbitrary
constants $A$ and $B$ to match the boundary conditions $\rho(0) = \rho_i$
and $\rho(L) = \rho_f$, and we see (figure~6) that $\rho$ is 
{\it expelled from the interior}: pulling it on the boundary only
affects a region of length $\sqrt{m}$, and the order parameter exponentially
decays into the bulk.  $\rho$ is constrained to zero in the inside of the
sample!

Why do we call this a mass?  The name comes from particle physics.  The photon
is massless.  Two charges $e_1$ and $e_2$ separated by a distance $r$
interact by a force whose magnitude goes as $e_1 e_2 / r^2$: this
is Coulomb's law.  The particle physicists interpret this force in
terms of the two particles exchanging ``virtual'' photons.  (I think
of the $1/r^2$ decay as the virtual photons being diluted over a sphere
of radius $r$.)  Now, the strong interaction between protons an neutrons 
has a different form: the force between them is always attractive, and goes as 
$e^{-\lambda r} / r^2$.  The exponential decay is extremely important,
since it keeps the nuclei of different atoms from attracting one another.
(We'd all have collapsed into neutron stars or worse were it not there!)
At long distances, the particle physicists interpret this force as
the proton and neutron exchanging virtual pions.\footnote{At
shorter distances, the picture is quarks exchanging gluons.  The
gluons have color, though, so the proton and neutron can't exchange
them at long distances.  Since colorless glueballs, if they exist,
are much more massive than pions, the dominant interaction for long
distances is pion exchange.}  Since the pion isn't
massless, the virtual pion field decays exponentially for exactly
the same reason that $\rho(x)$ decayed in our example above.

So, to enforce a constraint, we need to give the corresponding field a mass.
Let's see how that is done.

\section{The Meissner--Higgs Effect}

\begin{figure}[thb]
\epsfxsize=2.5truein
\epsffile{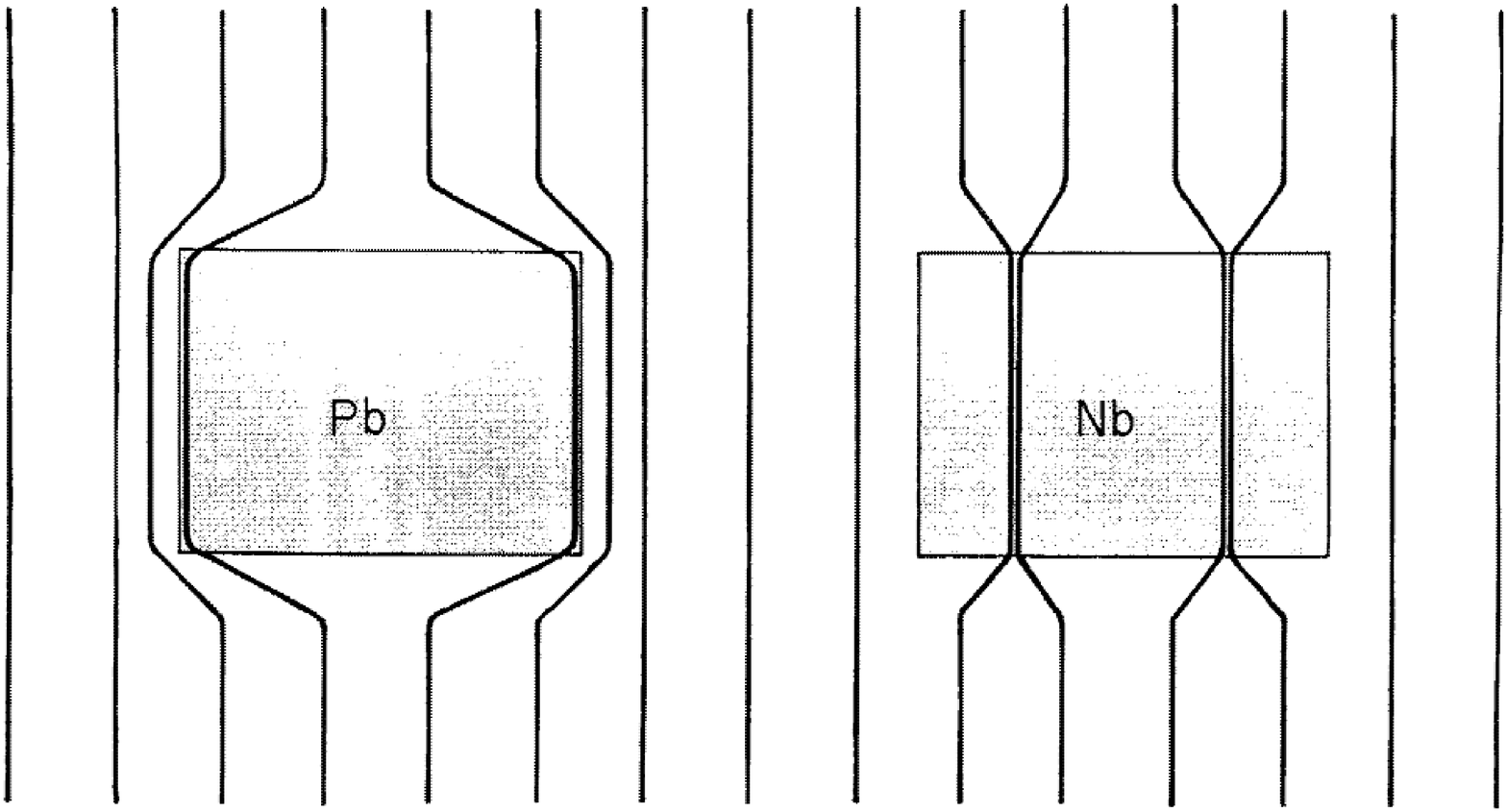}
\caption{{\bf Superconductors Expel Magnetic Fields.}
A magnetic field passing through a metal will be pushed out when the
metal is cooled through its superconducting transitions temperature.
This can happen in two different ways.  In type~I superconductors
like lead (chemical symbol Pb), the superconductivity is pushed entirely
outside the sample.  In type~II superconductors like niobium (Nb), the
magnetic field is broken up and confined to defect lines called vortices.
In both cases, the magnetic field is swept out of the remainder of the
sample.  The magnetic field penetrates a distance $\Lambda \sim 100$\AA
into the sample from the boundaries or from the vortex lines.
}
\label{fig:SuperconductorMeissner}
\end{figure}

In this section, I want to explain how superconductors expel magnetic
field.  This is a really beautiful argument, which I've basically
taken from Coleman's presentation.\cite{Coleman}  I'm afraid that there
is some math and a lot of physics that I need to introduce.  Most of
you will get lost, but the pictures will be nice anyhow.

\bigskip
\noindent
A. Introduction to the Meissner Effect.
\smallskip

Superconductors are named for their ability to carry currents of electricity
with absolutely no losses.  They have another, closely related property
which is no less amazing: they are a perfect shield for magnetic fields.
Remember the old science fiction stories about the scientist who finds
a material which is impervious to the gravitational field, paints
the bottom of his spacecraft with it, and falls to the moon?  Superconductors
work that way for magnetic fields.

Ashcroft and Mermin have a nice, not too technical discussion of
superconductors in one of the last chapters in their textbook.\cite{Ashcroft}
Figure~7 shows the two types of superconductors, represented by lead and
niobium.  At high temperatures, when the materials aren't superconducting,
the magnetic field penetrates the materials almost as if they weren't there.
(Iron would pull the magnetic field lines inward.)  Lead, when superconducting,
pushes the magnetic field out: just as for the example in section II,
the field a distance $r$ inward from the boundary decays like 
$B = B_0 e^{-r / \Lambda}$.
If you put too high a field, the lead will give up and let the field in:
but it will stop superconducting.  

On the right, we see that niobium behaves a bit differently.  It expels 
small magnetic fields like lead does, but larger fields are pushed into 
thin threads, called vortex lines.  These two general categories are
(rather unimaginatively) called type I and type II superconductors.
The vortex lines are the topological defects for the superconductor (lecture~1).
Superconductors are described by a complex number $\psi = \rho e^{i \theta} $,
whose magnitude $|\psi| = \rho$ is roughly constant.  The order
parameter at low temperatures is the phase $\theta$, and thus the
order parameter space is a circle ${\cal S}^1$.  A vortex line must
pass through any loop around which the phase of the order parameter changes by 
$2 \pi$.  The magnetic field in type~II superconductors decays like
$B = B_0 e^{-r / \Lambda}$ where here $r$ is the distance to the vortex
line.  Magnetic field is squeezed out of the bulk of the material into
these defects.

So, the magnetic field isn't actually stopped, it just peters out.
What kind of a leaky shield is that?  Actually, it's about as good
as one can hope: after all, the magnetic field won't be able to tell
it's in a superconductor until it gets inside a bit!  (Atoms don't go
superconducting, only huge heaps of atoms together, so the field has
to pass through a heap or two to realize that it isn't wanted.)  Anyhow,
$\Lambda$ is usually pretty small, a few hundred \AA ngstroms or so.
An 0.1mm thin layer of superconducting paint naively would let through a field
one part in $e^{-10000} \sim 10^{-4000}$ of the 
original.  Unfortunately, it usually doesn't work so well:
a few vortex lines get stuck on junk in the paint, and let in comparatively
large fields.

Before we can explain the repulsion of magnetic fields, we should explore
the broken symmetry.  Let's start with superfluids, which are simpler.

\bigskip
\noindent
B. Superfluid Free Energy and Spontaneous Symmetry Breaking.
\smallskip

\begin{figure}[thb]
\epsfxsize=2.5truein
\epsffile{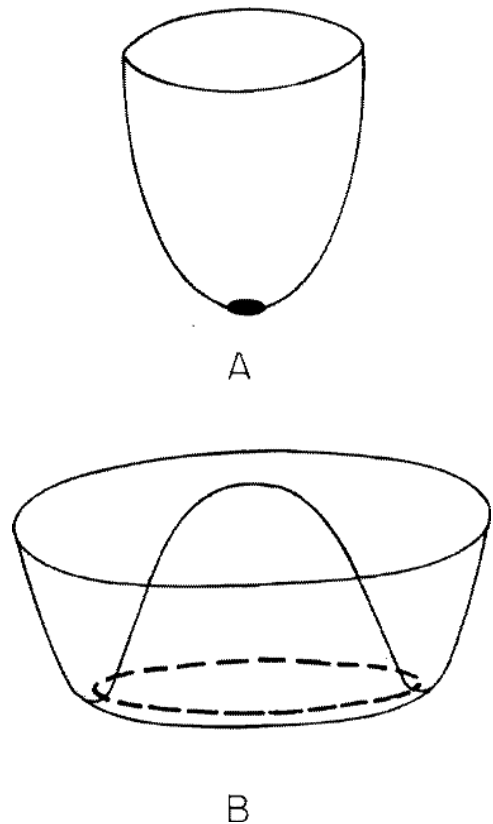}
\caption{
{\bf Superfluid Free Energy.} (a)~$T>T_c$: {\it Unbroken Symmetry.}
The free energy for a normal metal or fluid, above the 
superconducting or superfluid transition temperature, for a uniform
order parameter field $\psi$.  The vertical axis represents the
energy $\alpha |\psi|^2 + \beta |\psi|^4$, and the horizontal axes represent
the real and imaginary parts of $\psi$.  The coefficient $\alpha>0$,
so the minimum of the energy is at $\psi=0$.  Notice that the energy
is invariant under the symmetry $\psi \rightarrow e^{i \theta} \psi$
(corresponding to rotating the figure about the vertical).  This is
a symmetry of the free energy.  Notice also that the lowest energy state 
$\psi=0$ is also unchanged by this rotation: the symmetry is unbroken
above $T_c$.\hfill\break
(b)~$T<T_c$: {\it Broken Symmetry.}
The free energy ${\cal E}_{superfluid}$ for helium below the superfluid 
transition temperature.  The energy now looks like a Mexican hat: it
is still invariant under rotations about the vertical axis.
Since now $\alpha<0$, the energy is at a minimum along a circle, of radius 
$|\psi| = \sqrt{\alpha/2\beta}$ and arbitrary phase $\theta$.  The
superfluid must choose between these various possible phases, and
that choice breaks the symmetry.  This is a good example of spontaneous
symmetry breaking: just as the magnetization of a magnet selects
a direction in space and breaks rotational invariance, the superconductor
picks out a value of $\theta$.
}
\label{fig:MexicanHat}
\end{figure}

The order parameter for a superfluid, just as for a superconductor, is 
a complex number $\psi$.  The free energy for the superfluid is usually
written as\footnote{There are two new symbols here:
$\nabla = (\partial/\partial x, \partial/\partial y)$ and 
$|\chi|^2 = \chi^* \chi$, where $\chi^*$ is the complex conjugate of
$\chi$.  Written out in components,
${\cal E}_{superfluid} = \int dV 
(\partial \psi / \partial x)^* (\partial \psi / \partial x)
+(\partial \psi / \partial y)^* (\partial \psi / \partial y)
+ \alpha \psi^* \psi + \beta (\psi^* \psi)^2$
You can think of this as
a mathematical expression of the Mexican hat potential in figure 8b,
together with a resistance to abrupt changes in the order parameter.}
\begin{equation}
\label{eq:superfluid}
{\cal E}_{superfluid} = \int dV |\nabla \psi|^2 + \alpha |\psi|^2 + 
	\beta |\psi|^4.
\end{equation}
Above the superconducting transition temperature $T_c$, the coefficient
$\alpha > 0$.  If we imagine a constant order parameter field, the 
free energy forms a bowl (figure~8a) with a minimum at zero, as a function
of the real and imaginary part of $\psi$.  Zero order parameter corresponds
to a normal metal (for a superconductor), or a normal liquid (for a superfluid).

Below $T_c$, $\alpha<0$, and the potential is at a minimum for 
$\rho_0 = |\psi| = \sqrt{\alpha/2\beta}$: the potential in the complex plane
looks like a Mexican hat (figure~8b).  Now there are many possible ground
states: for any $\theta$, a constant order parameter field
$\psi = \rho_0 e^{i \theta}$ is a ground state.  Because the free energy
depends only on $|\psi|$ and  $|\nabla\psi|$, it is symmetric to changing
the phase $\theta$: the superconducting state chooses a specific value for
$\theta$, and thus {\it spontaneously breaks the symmetry}.  The 
circle of ground states in the brim of the Mexican hat is the order
parameter space for the superconductor.

We can write the free energy in terms of $\theta$:
\begin{equation}
\label{eq:Goldstone}
{\cal E}_{superfluid} = \int dV |\nabla \rho|^2 + \rho^2 |\nabla \theta|^2
+ \alpha \rho^2 + \beta \rho^4.
\end{equation}

As we discussed in the previous section, $\rho$ is ``massive''.
In figure~8b, if we vary $\rho$ slightly away from $\rho_0$, the energy
increases quadratically: $\alpha \rho^2 + \beta \rho^4 -
(\alpha \rho_0^2 + \beta \rho_0^4) \approx
        (\alpha + 6 \beta \rho_0^2) (\rho - \rho_0)^2$
The effective free energy for $\rho$ near $\rho_0$ is precisely
of the form \ref{eq:liquid} (except for unimportant constant shifts), with 
$m = \alpha + 6 \beta \rho_0^2$.  Thus just as before, $\rho$ will rapidly 
be drawn to its minimum energy state $\rho_0$.  Because $\rho$ is massive,
it is basically constrained to stay at its minimum value.  This is why it
is ignored at low temperatures in writing the order parameter field.
Here, the constraint doesn't do anything interesting: our next constraint
will be more interesting.

The $\theta$ field keeps the symmetry of the original model: rotating
it to $\theta + \theta_0$ doesn't change the energy a bit.  It is a 
Goldstone mode for our problem, and long--wavelength plane waves produce
what is known as ``second sound'' in superfluids.  Second sound turns
out to be heat waves: pulses of temperature which propogate like waves through
the superfluid.

\bigskip
\noindent
C. Superconducting Free Energy and the Higgs Mechanism
\smallskip

To describe the expulsion of magnetic field from superconductors, I have
to tell you how magnetic fields interact with the superconducting order.
I'm afraid this will be rather sketchy, and I apologize for trying.  

First of all, the particles which superconduct are pairs of electrons.
Electrons are charged, and repel one another with electric fields. 
Thus the electrons interact with electric fields.
We learn in the second semester of physics (if we're lucky) that electric
and magnetic fields are closely related to one another.  (This was
discovered by Einstein: a moving electric $E$ field develops a magnetic
$B$ component.)

Now, the $E$ and $B$ fields can be written at the same time in terms of
another field $A$.  It is this new field which is easiest to work with.
In particular, 
\begin{eqnarray}
\label{eq:B}
B &=& {\rm curl}\, A \\
&=& ({\partial A_z \over \partial y} - {\partial A_y \over \partial z},
~{\partial A_x \over \partial z} - {\partial A_z \over \partial x},
~{\partial A_y \over \partial x} - {\partial A_x \over \partial y}). \nonumber
\end{eqnarray}
The magnetic energy is ${\cal E}_{magnetic} = \int dV\,B^2$.

Now, you remember that I mentioned earlier that light (photons) is massless?
You may know that light is sometimes called ``electromagnetic radiation''.
The ``order parameter field'' for light is precisely the $A$ field.
We can see by expanding $B^2$ in terms of $A$ that the energy for the $A$ field
\begin{equation}
\label{eq:light}
{\cal E}_{magnetic} = 
	\int dV (\partial A_z/\partial y - \partial A_y/\partial z)^2 + \cdots
\end{equation}
doesn't have any terms like $A^2$.  When we add the energy from the electric
fields, this is still true: light is massless because the electromagnetic
energy involves only derivatives of $A$.

Now, I need to know how the electromagnetic order parameter $A$ interacts
with the superconducting order parameter $\psi$.  I'll just tell you.
The free energy for a superconductor looks like
\begin{equation}
\label{eq:superconductor}
{\cal E}_{superconductor} = \int dV\,|\nabla \psi - i A \psi|^2 
+ \alpha |\psi|^2 + \beta |\psi|^4 + B^2
\end{equation}
If we set $\psi=0$, we get the magnetic energy $B^2$ for the $A$-field.
If we set $A=0$, we get the superfluid energy \ref{eq:superfluid}.  I don't
know of a way to motivate the way in which we couple the $A$ field
to the gradient $\nabla \psi$.  I don't think anyone has
a simple derivation.  This way of connecting the two is called ``minimal
coupling'', which just gives a name to the unexplained fact that the
simplest way of coupling the two gives the right answer.

Now, if we assume $T < T_c$, so $\alpha<0$ and $\rho \sim \rho_0 e^{i\theta}$,
we find
\begin{equation}
\label{eq:sc2}
{\cal E} \approx \int dV\, \rho_0^2\, |\nabla \theta - A|^2 + 
(\partial A_z/\partial y - \partial A_y /\partial z)^2 + \cdots.
\end{equation}
We want to know if $A$ or $\theta$ is going to develop a mass.  The
problem is, ${\cal E}_{superconductor}$ doesn't look quite like the form 
\ref{eq:liquid} for either one.  If we combine the two into a new order parameter 
field
$C = \nabla \theta - A$, and use the fact that the second partial derivative
$\partial^2 \theta/\partial z \partial y =
		\partial^2 \theta/\partial y\partial z$,
we see that 
\begin{eqnarray}
{\rm curl}\, C 
	 &=& (\partial C_z/\partial y-\partial C_y/\partial z,~\cdots)\nonumber\\
         &=& (\partial A_z/\partial y - \partial A_y /\partial z,~\cdots)\nonumber\\
	 &=& B
\end{eqnarray}
so 
\begin{equation}
\label{eq:sc3}
{\cal E} \approx \int dV\, \rho_0^2\, C^2 + 
(\partial C_z/\partial y - \partial C_y /\partial z)^2 + \cdots.
\end{equation}
Thus the new, combined field $C$ is massive.  $C$ will be constrained
to zero in the bulk, exponentially decaying like $C_0 e^{-\rho_0 r}$.
The magnetic field $B = {\rm curl}\,C$ thus also decays, and the 
penetration depth $\Lambda = 1/\rho_0$.

We started with a massless photon field $A$ and a massless Goldstone mode 
$\theta$.  We ended up with only one field $C$, with a mass.  Did we lose
something?  No, actually $C$ has three components: two components 
corresponding to the original two polarizations of light, and one component
corresponding to the Goldstone mode.  Coleman\cite{Coleman}
says ``the Goldstone boson eats the photon, and gains a mass''!

The Weinberg--Salaam theory of the weak interaction is exactly analogous
to the theory of superconductivity.  The role of lead or niobium is played
by the vacuum.  The free energy of the universe has an $SU(3)$ symmetry,
which is spontaneously broken to a smaller symmetry $SU(2)\times U(1)$.
The $W^\pm$ and $Z$ bosons which now mediate the weak interaction used
to be massless: they and the photon were all part of one big $A$-field.
If current theories of cosmology are true, this ``superconducting''
transition occurred in the first instants after the Big Bang.

Now, after explaining superconductors, the weak interaction, and the 
phase transition in the early universe, let's return to why we don't 
understand crystals.

\section{The Mystery of the Crystals}

\begin{figure}[thb]
\epsfxsize=2.5truein
\epsffile{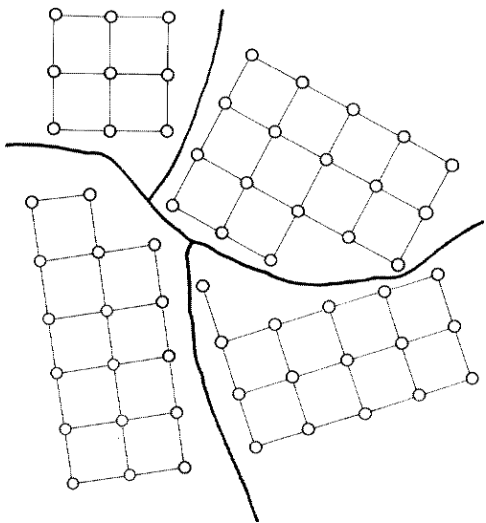}
\caption{
{\bf Polycrystal.}
Many crystalline materials, such as metals, normally aren't made of a single
crystal.  They are formed from many crystalline domains: a polycrystalline
configuration.  I show a schematic of a polycrystal here.  The important
thing to notice is that the atoms within a domain are almost undeformed
except right next to the domain wall.  All the rotational deformation is
expelled into sharp domain boundaries.
}
\label{fig:Polycrystal}
\end{figure}

Normally, when you think of crystals, you think of diamonds, snowflakes,
or maybe salt crystals.\footnote{Some of you will think of wine 
glasses.  They are made of glass and aren't crystals at all.}  These are 
single crystals: the sodium and chlorine atoms in a grain of salt sit
in registry all the way across the grain, giving it its cubical shape.
Did you know that metals form crystals?  In the last lecture, I 
mentioned dislocation lines in a copper crystal.  Metals don't have
big facets and corners because they are polycrystalline.  The atoms
in a metal also sit on a regular lattice, but the metal breaks up
into domains in which the lattices sit at various angles (figure~9). 
Because there are lots of small domains, copper doesn't form 
facets like salt grains and snowflakes do.\footnote{Metal crystals
are sometimes found in nature.  The growth takes place so slowly that a
single crystal can form.  The same idea happens with rock candy: you
get a glass if you cool sugar syrup quickly, but if you evaporate
a sugar solution slowly, you can get big crystals.}

\begin{figure}[thb]
\epsfxsize=2.5truein
\epsffile{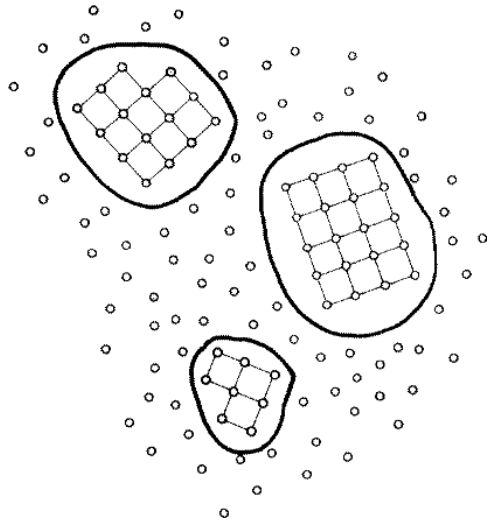}
\caption{{\bf Growing a crystal from a liquid: forming a polycrystal.}
Polycrystals can form for lots of reasons.  If one cools a liquid quickly,
one can find that crystalline regions can form in many different places
almost simultaneously.  Since they will have random orientations, they
won't match up when they meet.  When they do meet, rearrangements of
atoms will occur to try to realign and merge the domains (coarsening).
As we continue to cool and wait, this process will eventually stop,
leaving us with different domains.
}
\label{fig:GrowingCrystal}
\end{figure}

What Ming Huang (one of my students\cite{Ming}) and I have been trying 
to explain for
years is why those little domains form.  It's easy to see that different
regions might grow with different orientations (figure~10).  When they touch, 
the different domains will start pushing and twisting one another, trying to 
make one big domain.  It isn't hard to believe that they will stop growing
after a while, fighting one another to a standstill.  What we've been
trying to understand, though, is why the final state is made of perfect
little crystals separated by sharp domain walls.

Now, I don't want to exaggerate.  There are perfectly good explanations
for why crystals form domain walls.  They just aren't as beautiful and
general as they might be.  They don't fit in with the general ideas
of broken symmetries and order parameters: they apply only to crystals.
Our explanation for why superconducters don't
have a Goldstone mode was perfectly OK before Higgs came too.  He made
it beautiful and generalized it to explain something completely different.
Ming and I want to understand grain boundaries in a way which will make
simple and clear where else similar phenomena might occur.  At least,
we'd like to understand why focal conics occur at the same time.  Domains
formed by breaking translational symmetry in one direction and in three 
directions should have the same kind of explanation!

\begin{figure}[thb]
\epsfxsize=2.5truein
\epsffile{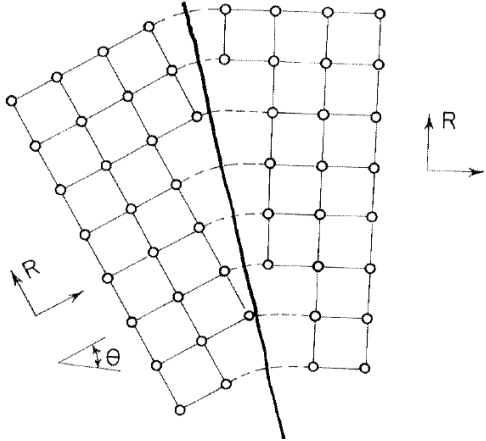}
\caption{{\bf Domain Wall.}
Here we see a single domain wall.  Notice that the domain wall can
be also thought of as a series of dislocations.  The strain field
inside the crystal due to a line of dislocations can be shown to
decay exponentially, just as the magnetic field dies away around
a vortex line.
}
\label{fig:GrainBoundary}
\end{figure}

Figure~11 shows a domain wall in a crystal.  The crystalline ground state
rotates as one crosses the domain wall.  The atoms at the wall are
quite unhappy.  You'd think that they would push and pull on their 
neighbors, and that there would be strains leaking far into the crystal.
This isn't true.  In fact, there is a well--known rule in the materials
science literature, that the strain field from a domain wall dies away
exponentially as one enters the grain.

Doesn't that sound like a Meissner effect?  

\begin{figure}[thb]
\epsfxsize=2.5truein
\epsffile{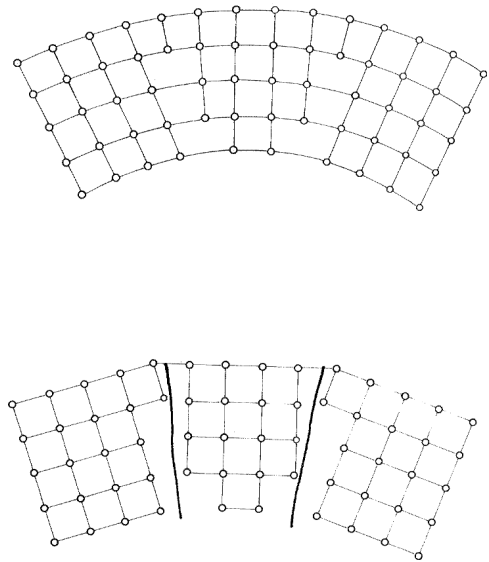}
\caption{
{\bf (a)~Rotational distortion of a crystal.}
If we take a thick piece of metal and rotate one end with respect to another,
it will start by bending uniformly.  As it continues to bend, 
dislocations will form to ease the bending strain.  These line dislocations
will start off distributed irregularly through the sample.
{\bf (b)~Domain walls form to expel rotations.}
If we hold the rotation for a long time, and let the dislocations move
around, they will lower their energy by arranging themselves into domain 
walls.  Between the domain walls we find undistorted crystal.  This
process is called polygonalization.
}
\label{fig:Polygonalization}
\end{figure}

There are more analogies.  Crystals break both the translational and
the rotational symmetry of liquids.  Many liquid crystals only
break the rotational symmetry.  They have Goldstone rotational waves:
if you rotate a large region inside a liquid crystal, it will cost
little energy, and will slowly rotate back.  When the translational
symmetry is also broken, the rotational Goldstone mode disappears!
If I rotate one piece of a crystal with respect to another, it costs an 
enormous energy (figure~12a).  If I let the distorted crystal rearrange
locally to reach equilibrium, the rotational deformation
is expelled into grain boundaries (figure~12b), a process known in
the field as polygonalization.  Just like the massless photon developed
a mass when the superconducting transition broke the gauge symmetry,
the massless rotational mode develops a mass when the translational
symmetry is broken.

This is surely also related to some of the old problems in the topological
theory of defects.  In describing a crystal, everybody uses the displacement
field $u(x)$ and its derivatives.  Now, as we saw in lecture~1, 
$u(x)$ describes the broken translational order, but not the broken
orientational order.  Why don't we also have a rotation matrix $R(x)$?
For example, in figure~11, $R(x)$ shifts abruptly from one side of the
domain wall to the other.  Mermin\cite{Mermin} discusses some of the
weird behavior one gets following this path.  The point is, $R(x)$ seems to
be constrained: it doesn't change on its own, but follows the broken 
translational order.  Keeping it as an order parameter seems no more 
necessary than keeping $\rho = |\psi|$ around in a superconductor: only 
$\theta$ is massless, and $\rho$ just wiggles around $\rho_0$ in a 
boring way.

Now, Ming and I have spent a huge amount of time trying to make these
words into a mathematical theory.  (We started with smectics, then
studied superconductors, then thought about some ideas of Toner and Nelson,
$\dots$)  Ming has gone on to better things, and I'm still futzing with
it.  I can summarize where we are right now.  Suppose we consider
a rotationally distorted two--dimensional crystal (figure~12a).  We can 
define a rotational order parameter by looking at the angle of the
nearest--neighbor bonds:
\begin{equation}
\label{eq:R}
R(x) = \left(\matrix{\cos\theta&\sin\theta\cr 
			     -\sin\theta&\cos\theta\cr}\right).
\end{equation}
The translational order parameter $\vec u$ is just as it always was: if 
$\vec x$ is the original position and $\vec p(x)$ is the corresponding 
position in the ideal lattice,
\begin{equation}
\label{eq:u}
\vec u(x) = \vec p(x) - \vec x.
\end{equation}
Now, the free energy can only depend on gradients of $\vec u$, since
it is translationally invariant.  It also cannot change if we perform
a uniform rotation: $R \to R_0 R$, $p \to R_0 p$.  From this, we
can see that the free energy must be written in terms of gradients
of $R(x)$ and the particular combination\footnote{This
is analogous to the minimal coupling term $\nabla \theta - A$ in the
free energy for a superconductor.}
\begin{equation}
\label{eq:covariant}
\epsilon_{ij} = \delta_{ij} 
	- \sum_{k=1}^2 R_{ki} (\partial u_j/\partial x_k + \delta_{kj}).
\end{equation}
A reasonable free energy for a crystal then becomes
\begin{eqnarray}
\label{eq:crystal}
{\cal E}_{crystal} &=& (\nabla \theta)^2 
	 + 2 \mu \sum_{ij} \left({\epsilon_{ij}+\epsilon_{ji}\over 2}\right)^2 \\
		&+& \lambda \sum_{i} \epsilon_{ii}^2 
		+ \kappa \left({\epsilon_{12}-\epsilon_{21}\over 2}\right)^2.
			\nonumber
\end{eqnarray}
This is just the normal elastic energy everybody uses, except for the third
term multiplied by $\kappa$.  Normally, the strain matrix $\epsilon$ is
defined to be symmetric, so this term is then zero.

Our free energy doesn't keep $\epsilon$ automatically symmetric precisely
because we have $R(x)$ as an independent degree of freedom.  The antisymmetric
part measures the amount that $R$ disagrees with the local gradients of 
$\vec u$.  It turns out that this antisymmetric part for the crystalline
free energy is analogous to the current for the superconductor, which
has a Meissner effect.\footnote{It is the gradient of 
${\cal E}_{crystal}$ with respect to $\theta$, just as the current
is the gradient of ${\cal E}_{superconductor}$ with respect to $A$.
I thank Alan Luther for pointing this out.}  

There are several things I haven't been able to do, though.  First, I don't
think $\epsilon_{12}-\epsilon_{21}$ is expelled quite like its analogue
in the superconductor.  I think we can show, though, that it is a 
boring variable like $\rho$ was.  Second, I haven't a clue on how to show 
that grains exist.  To show that grains exist I have to show a constraint
like $\nabla \theta = 0$!

\bigskip

We started this lecture by admiring the focal conic defects in smectic
liquid crystals: beautiful ellipses and hyperbolas which are due not
to topology, but to geometrical consequences of a constraint.  We saw
how constraints can be enforced by the energy: ``massive'' modes decay 
exponentially.  We saw explicitly how this occurs in superconductors ---
the magnetic field is constrained to zero because the photon and the
Goldstone boson for the superconducting gauge symmetry combine into
a massive particle.  Finally, we discussed analogous effects in the
everyday problem of grain boundaries in crystals, and realized that
we don't really understand them in a deep sense.

\section{Appendix: The Smectic Order Parameter}

\begin{figure}[thb]
\epsfxsize=2.5truein
\epsffile{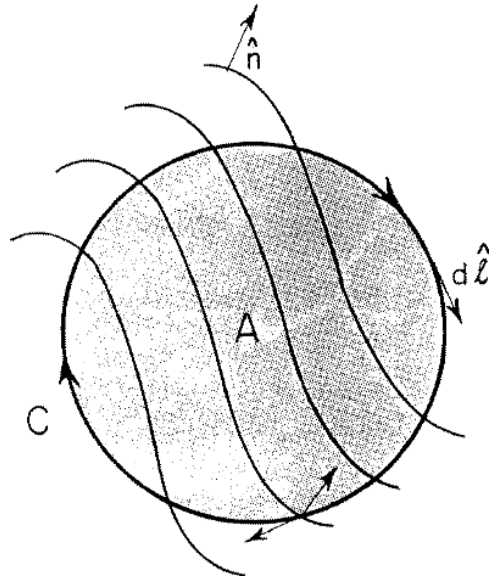}
\caption{
{\bf Equally Spaced Layers Imply ${\rm curl}\, n = 0$.}
Smectic layers, with a loop $C$ enclosing an area $A$.  The dot product
$\hat n \cdot d\ell$ gives the cosine of the angle of the curve $C$ with 
respect to the layers, and $a/\cos\theta$ is the length of curve $C$ between
two layers, so $1/a \int_C \hat n \cdot d\ell$ gives the net number of layers 
crossed by the curve $C$.  (A layer crossed first forward and then backward
cancels, of course).  Since in a closed loop the net number of layers
crossed must be zero (assuming no dislocations), this must be zero.
By Stokes' theorem, $\int_C \hat n \cdot d\ell = \int_A {\rm curl}\,n \cdot dA$.
This is true for any little area $A$, so ${\rm curl}\,n \equiv 0$.
}
\label{fig:CurlN0}
\end{figure}

Here we derive the consequences for layered systems of the constraint
that the layers be equally spaced.  Suppose that there are a stack
of (bent) sheets, equally spaced from one to the next, with separation $a$.
Suppose that the unit normal to these sheets at a position $\vec x$ is
given by $\hat n$.  Consider traveling around a loop $C$, crossing various
layers as we go around (figure~A1).  The number of layers we cross is
given by the line integral
\begin{equation}
\label{eq:crossed}
(1/a) \int_C \hat n \cdot d\ell = {\rm net~\#~crossed}.
\end{equation}
If the layers exist throughout the region without any defects, then
the net number crossed around any closed loop must be zero.  Using
Stokes' theorem, this integral over $C$ is equal to an integral over
the area $A$ swept out by the curve:
\begin{equation}
\label{eq:Stokes}
\int_C \hat n\cdot d\ell = \int_A {\rm curl}\,\hat n\cdot dA.
\end{equation}
But for this to be true for all areas $A$, ${\rm curl}\, \hat n$ must be zero.

Now, we already know that $\hat n^2 = 1$.  The derivative
$\partial \hat n^2/\partial x_\alpha$, of course, must be zero, so using 
the product rule
\begin{equation}
\label{eq:n}
\sum_\beta \hat n_\beta\, \partial \hat n_\beta/\partial x_\alpha = 0.
\end{equation}
Now, since we know ${\rm curl}\, \hat n = 0$, we know from \ref{eq:curl} that
\begin{equation}
\label{eq:equal}
\partial\hat n_\beta/\partial x_\alpha
	=\partial\hat n_\alpha/\partial x_\beta.
\end{equation}
Finally, combining these, we find
\begin{equation}
\label{eq:generator}
\sum_\beta \hat n_\beta\, \partial \hat n_\alpha/\partial x_\beta = 
	(\hat n \cdot \nabla)\, \hat n = 0.
\end{equation}
This implies that $\hat n$ doesn't change when you move in the $\hat n$ 
direction.  This means that $\hat n$ will be the perpendicular to 
the next layer as well: that is, a straight line perpendicular to one
layer will be perpendicular to every layer it crosses.  

These perpendicular lines are called generators.  We qualitatively knew
already that one layer determined its surroundings: now we have a simple
geometrical rule describing this nonlocal constraint.  For your information,
the defects occur where the generators cross (as shown in figure~3):
this surface is called the evolute, or surface of centers, for the
layer.

\bigskip\bigskip
\centerline{\bf Acknowledgments}
\medskip

I'd like to acknowledge NSF grant \# DMR-9118065, and thank NORDITA and the
Technical University of Denmark for their hospitality while these
lectures were written up.

\end{document}